\documentclass[11pt, a4paper]{article}
\usepackage[english]{babel}
\usepackage{amsfonts,amsmath,amsthm,amssymb}
\usepackage[margin=1.2in]{geometry}
\usepackage[utf8]{inputenc}
\usepackage{float}
\usepackage{epsfig}
\usepackage{graphicx}
\usepackage{cancel}
\usepackage{hyperref}
\usepackage{mathrsfs} 
\usepackage{bm} 
\usepackage{tikz}
\usepackage{upgreek}
\usetikzlibrary{arrows,positioning,cd,calc,math,decorations.pathreplacing,decorations.markings,decorations.pathmorphing,patterns
}
\tikzset{wave/.style={decorate, decoration=snake}}
\usepackage{bbold}
\usepackage[backend=biber,
style=alphabetic, maxnames=50]{biblatex}
\providecommand*\email[1]{\href{mailto:#1}{#1}}

\usepackage{subcaption}
\numberwithin{equation}{section}
\usepackage{xcolor}
\definecolor{MyGreen}{rgb}{0.05, 0.7, 0.06}
\definecolor{MyBlue}{rgb}{0.06, 0.3, 0.57}
\definecolor{MyRed}{rgb}{0.9,0.19,0.39}
\definecolor{MyOr}{rgb}{1.0, 0.55, 0.0}
\definecolor{darkspringgreen}{rgb}{0.05, 0.5, 0.06}

\newcommand{\tr}{\operatorname{tr}}

\newtheorem{theorem}{Theorem}%
\newtheorem{proposition}[]{Proposition}
\newtheorem{definition}{Definition}
\newenvironment{sproof}{%
  \proof}{\endproof}
\newtheorem{remark}{Remark}
\newcommand{\eq}[1]{\begin{equation}\begin{gathered}#1\end{gathered}\end{equation}}
\newcommand{\eqs}[1]{\begin{equation}\begin{gathered}\begin{split}#1\end{split}\end{gathered}\end{equation}}

\DeclareMathOperator{\res}{Res}
\newcommand{\sfa}{{\sf a}}
\newcommand{\Tau}{{\mathcal{T}}}
\newcommand{\sfb}{{\sf b}}
\newcommand{\sfc}{{\sf c}}
\newcommand{\sfd}{{\sf d}}
\newcommand{\sfQ}{{\sf Q}}
\newcommand{\sfY}{{\sf Y}}
\newcommand{\mc}{\mathcal}

\newcommand{\cP}{\mathcal{P}}
\newcommand{\cin}{\mathcal{C}_{in}}
\newcommand{\cout}{\mathcal{C}_{out}}
\newcommand{\cC}{\mathcal{C}}
\newcommand{\cH}{\mathcal{H}}
\newcommand{\Yt}{\widetilde{Y}}

\newcommand{\T}{\mathcal{T}}

\newcommand{\diag}{\text{diag}}

 \date {}

\begin{document}

\title{Painlev\'e/CFT correspondence on a torus}
\author{Harini Desiraju  \\
{\email{harini.desiraju@sydney.edu.au}}
  \\
  \normalsize\it School of Mathematics and Statistics F07, The University of Sydney, NSW 2006, Australia,}
\maketitle

\begin{abstract}
This note details the relationship between the isomonodromic tau-function and conformal blocks, on a torus with one simple pole. It is based on the author's talk at ICMP 2021.
\end{abstract}
 
\tableofcontents
\pagebreak

The seminal paper \cite{Gamayun:2012ma} by Gamayun, Iorgov, Lisovyy introduced the relation between the Painlev\'e VI equation and conformal field theory (CFT) on a sphere with four punctures. Since then, such a correspondence has been extended to include other Painlev\'e equations \cite{Gamayun:2013auu, bonelli2017painleve}. A rigorous approach to the Painlev\'e/CFT correspondence comes through the formulation of isomonodromic tau-functions as Fredholm determinants \cite{GL, cafassotau, gavrylenko2018pure}. In a recent paper \cite{del2020isomonodromic} we rigorously proved the Painlev\'e/CFT correspondence on the torus with any number of simple poles, using Fredholm determinants. This note provides a detailed treatment of the Painlev\'e/CFT correspondence for the case of a one-point torus. 

\section{A quick overview of the story on a Riemann sphere}

Painlev\'e equations were first formulated in the 1900s by Fuchs \cite{fuchs1905quelques}, Gambier \cite{gambier1910equations}, Painlev\'e \cite{painleve1906equations}, and Picard \cite{picard1889memoire} in their quest to find second order differential equations whose solutions are such that their only movable singularities are poles. There are 6 such equations and they take the form
\begin{gather}
\frac{d^2 u}{dt^2} = F\left( u, \frac{du}{dt}, t \right) \label{Painleve}
\end{gather}
with $F$ rational in $u, \frac{du}{dt}$, analytic in $t$. The solutions of these equations are transcendental in nature and can be viewed as nonlinear analogues to special functions. A century later, these equations found a renewed purpose in mathematical physics due to the discovery \cite{FN, garnier1912equations, JMU1} that the Painlev\'e equations are integrable, i.e they have an associated Lax pair 
\begin{gather}
    \frac{\partial Y(z,t)}{\partial z} =  Y(z,t)L_{1}(z,t) , \qquad \frac{\partial Y(z,t)}{\partial t} =  Y(z,t) L_{2}(z,t), \label{eq:Lax_pair}
\end{gather}where the matrices $L_{1}, L_{2}$ are meromorphic matrix valued functions on an $n$-puntured Riemann sphere $C_{0,n}$ depending on $z, t, u(t), d u(t)/dt$, and the consistency condition  
\begin{gather}
    \frac{\partial^2 Y(z,t)}{ \partial t \partial z} =  \frac{\partial^2 Y(z,t)}{ \partial t \partial z}  \Rightarrow \frac{\partial L_{1}}{\partial t} - \frac{\partial L_{2}}{\partial z} - [L_{1}, L_{2}] =0 \label{def:zero_curv}
\end{gather}
gives back the Painlev\'e equation \eqref{Painleve}. Moreover, the solution $Y(z,t)$ is multivalued with the monodromy data (Stokes data included) dictated by the order of the singularities on the Riemann sphere $C_{0,n}$. The Painlev\'e equation therefore dictates the monodromy preserving deformations of the system \eqref{eq:Lax_pair}, and in this sense, the variable $t$ is called the {\it isomonodromic time} variable.

Painlev\'e equations also have an associated Hamiltonian structure \cite{JMU1, okamoto1986studies, forrester2001application} \footnote{See \cite{NIST_DLMF} eq. 32.6 for a comprehensive list}
\begin{gather}
    H := \frac{1}{2}\res \tr L_{1}^2 \label{def:Ham}
\end{gather}
such that \eqref{Painleve} are the equations of motion of the Hamiltonian above. Here $\res$ means the residue at all the singularities in $z$ of the Lax matrix $A(z,t)$.

A central object in the study of integrable systems is the so-called tau-function, which not only generates the Hamiltonians governing its dynamics, but also contains important information, in the present case, about the Painlev\'e transcendent such as the singular locus, and the asymptotic data. Refer to \cite{harnad2021tau} for a thorough treatment of the subject.
\begin{definition} Isomonodromic tau-function can be formulated as a generator of the Hamiltonian \eqref{def:Ham} as \cite{JMU1, its2018some}
\begin{gather}
    \partial_{t} \log \Tau := H. \label{tau_ham_def}
\end{gather}
\end{definition}
The solutions of Painlev\'e equations $u(x)$, known as the Painlev\'e transcendents, are then typically described by the logarithmic derivatives of the isomonodromic tau-function. For example, 
for Painlev\'e II, 
\begin{gather}\label{eq:solution_tau}
    u_{PII}(t)^2 = - \frac{d^2}{d t^2} \log \Tau,
\end{gather}
and for the other Painlev\'e equations, the analogue of the above relations may involve some factors of $t$.
Due to the Riemann-Hilbert correspondence, which provides a map between the initial data of the Painlev\'e equation and the associated monodromy data obtained through the solution $Y(z,t)$ of the system \eqref{linear_system}, the tau-function can be formulated in terms of the monodromy data  \cite{Mal, MalB}. The zeros of such tau-function determine where the Riemann-Hilbert map ceases to be bijective. Furthermore, as a consequence of the relations such as \eqref{eq:solution_tau}, the Painlev\'e transcendents can be expressed as a function of the monodromy data and isomonodromic time. Such an expression is key to understanding the relation to conformal blocks as will be shown in this note for the genus 1 case.

In late 1970s Jimbo, Miwa, Sato (dubbed the Kyoto school) formulated the tau-function as a 2-point correlation function of what they called holonomic fields in a series of papers starting from \cite{satoHol1}, in a first step towards giving a field theoretic interpretation of Painlev\'e equations. This program came to full fruition for the Painlev\'e equations III, V, VI, only in mid 2010s with the papers of Gamayun, Iorgov, Lisovyy \cite{Gamayun:2012ma, Gamayun:2013auu} that established a connection between some Painlev\'e equations and 2 dimensional CFTs, and was more rigorously proven in \cite{cafassotau, GL}. The crucial technique in \cite{GL, cafassotau} was that the tau-function of three (III, V, VI) of the six (I-VI) Painlev\'e equations could be written as a Fredholm determinant of some operator \cite{GL, cafassotau}, and its minor expansion is a combinatorial expression in terms of some Virasoro conformal blocks $\mathcal{B}$,
\begin{gather}
        \Tau= \sum_{n \in\mathbb{Z}} e^{\pi i n \eta } \mathcal{B}(\sigma+ n, t).  \label{tau:fourier}
    \end{gather}
For example, the above relation states that the tau-function of the Painlev\'e VI equation can be expressed in terms of Virasoro conformal blocks $\mathcal{B}$ of a CFT with central charge $c=1$ that are explicitly determined by hypergeometric functions. The expression \eqref{tau:fourier} can equivalently,  due to the Alday-Gaiotto-Tachikawa (AGT) correspondence \cite{Alday:2009aq}, be written in terms of Nekrasov-Okounkov partition functions of a supersymmetric gauge theory with so called self-dual ($\epsilon_{1}+\epsilon_{2}=0$) omega background, which are combinatorial sums over random partitions. The conformal block expansions of the Painlev\'e equations I, II, IV are conjectured in \cite{bonelli2017painleve} The Fredholm determinant representation for Painlev\'e II equation was formulated by the author in \cite{desiraju2019tau, desiraju2021fredholm}. 
A few comments are in order:
\begin{itemize}
    \item[1.] the expression of tau-functions as Fredholm determinants and as conformal blocks is remarkable due to the transcendental nature of the solutions,
    \item[2.] the ratio of Painlev\'e transcendents at the critical points, known as the connection constant is related to the fusion kernel of the associated conformal blocks,
    \item[3.] the zero locus of the Fredholm determinant uniquely determines the poles of the transcendent, also known as the Malgrange divisor. 
\end{itemize} 

With all of the above discussion in mind, we detail the extension of the relationship between Painlev\'e type equations and CFTs on the torus with one Fuchsian singularity $C_{1,1}$. It will be shown that the isomonodromic tau-function of a special case of the elliptic Painlev\'e VI equation can be written, first as a Fredholm determinant, and then as a Fourier series of conformal blocks. The discussion in the following section holds in general for a torus with any number of Fuchsian singularities and the corresponding Lax matrices that are $N\times N$, as shown in the paper of the author along with P. Gavrylenko, F. Del Monte \cite{del2020isomonodromic}. 

\subsection*{Acknowledgements}
The author thanks F. Del Monte, P. Gavrylenko, for useful discussions and suggestions.

\section{Painlev\'e/CFT correspondence on the one-point torus}



 Our goal in the paper \cite{del2020isomonodromic} was two fold:
\begin{itemize}
    \item[1.] to explicitly formulate isomonodromic tau-functions on a torus and thereby find a way to understand the solutions of isomonodromic equations on a torus explicitly, 
    \item[2.] to make express the isomonodromic tau-functions in terms of Nekrasov-Okounkov functions or equivalently conformal blocks.
\end{itemize}

We now demonstrate the above points using the simplest example of a torus with one simple pole:  consider a special case of the elliptic Painlev\'e VI equation: the equation of motion of the 2-particle non-autonomous Calogero-Moser system (in the center of mass reference frame),
\begin{align}\label{eq:EllPainleve}
    (2\pi i)^2\frac{d^2Q(\tau)}{d\tau^2}=m^2\wp'(2Q(\tau)|\tau).
\end{align}
The modular parameter $\tau$ assumes the role of the isomonodromic time and brings in an additional factor of $2\pi i$ w.r.t to the sphere case in the introduction. 
The above equation \eqref{eq:EllPainleve} is integrable in the sense that it has an associated Lax pair
\begin{equation}
\begin{split}
   \partial_z \mathcal{Y}_{CM}(z,\tau)= \mathcal{Y}_{CM}(z,\tau) L_{1}(z,\tau); \quad
    2\pi i \partial_\tau \mathcal{Y}_{CM}(z,\tau)=  \mathcal{Y}_{CM}(z,\tau) L_{2}(z, \tau) ,
    \end{split}\label{linear_system}
\end{equation}
where
\begin{gather}
    L_{1}(z, \tau)=2\pi i\frac{dQ(\tau)}{d\tau} \sigma_{3} + {m}\frac{\theta_1'(0\vert \tau)}{\theta_1(z\vert \tau)} \left(\begin{array}{cc}
        0 & \frac{\theta_1(z+2Q(\tau) \vert \tau)}{\theta_1(-2Q(\tau) \vert \tau)} \\
        \frac{\theta_1(z-2Q(\tau) \vert \tau)}{\theta_1(2Q(\tau) \vert \tau)} & 0
    \end{array}\right), \label{eq:Lax_A} \\
    L_{2}(z, \tau)=\left(\begin{array}{cc} 
        0 &m y(-2Q(\tau),z) \\
        my(2Q(\tau),z) & 0
    \end{array}\right), \label{eq:Lax_B}
\end{gather}
and the zero curvature equation \eqref{def:zero_curv}
gives back \eqref{eq:EllPainleve}. The notation for the solution $\mathcal{Y}_{CM}$ as opposed to $Y_{CM}$ will be addressed in definition \ref{def:YcalY}.
The toric analogue to the Hamiltonian \eqref{def:Ham} of the system above is the following a-cycle integral
\begin{equation} \label{eq:Ham_CM}
    H(\tau)=\oint_{a} dz\frac{1}{2} \tr L_{1}^2(z, \tau)=\left( 2\pi i\frac{dQ(\tau)}{d\tau} \right)^2-{m}^2\wp(2Q(\tau)|\tau)+4\pi i{m}^2\partial_\tau\log\eta(\tau),
\end{equation}
where $\eta(\tau)$ is Dedekind's eta function
\begin{equation*}
    \eta(\tau):=\left(\frac{\theta_1'(0\vert \tau)}{2\pi} \right)^{1/3}.
\end{equation*}
The tau-function is therefore
\begin{gather}\label{eq:IsomTau11}
    2\pi i \partial_{\tau} \log \T_{CM}(\tau) := H(\tau).
\end{gather} 
In the form above, the tau-function depends on the initial data of the equation \eqref{eq:EllPainleve}. Expressing the tau-function in terms of the monodromy data of the problem reveals the connection to conformal blocks as we show in subsequent sections.

\subsection{Monodromy problem}
 The Lax matrix \eqref{eq:Lax_A} has a simple pole at $z=0$ and is multivalued:
\begin{align}\label{L_1}
L_{1}(z+1,\tau)=L_{1}(z, \tau), &&  L_{1}(z+\tau, \tau)=e^{-2\pi iQ(\tau)\sigma_3} L_{1}(z, \tau) e^{2\pi iQ(\tau)\sigma_3}.
\end{align}
In turn, the monodromies around the A, B-cycles, and the puncture at $z=0$ read
\begin{gather}
\begin{array}{c}
 \mathcal{Y}(z+1, \tau)= M_A \mathcal{Y}(z, \tau)  , \qquad \mathcal{Y}(z+\tau, \tau)=   M_B  \mathcal{Y}(z,\tau) e^{2\pi iQ(\tau)\sigma_3}, \\ \\
\mathcal{Y}(e^{2\pi i}z, \tau)= M_0 \mathcal{Y}(z, \tau),
\end{array}
\end{gather}
and the monodromy matrices $M_{A}, M_{B}, M_{0}$ satisfy the constraint
\begin{equation}
    M_0=M_A^{-1} M_B^{-1} M_A M_B. \label{eq:monodromy_constraint}
\end{equation}
Explicitly, the monodromies around the A-cycle, and the puncture $z=0$ can be imposed to be
\begin{gather}
    M_A=e^{2\pi i {a} \sigma_3}, \qquad M_0\sim e^{2\pi i{m}\sigma_3}, 
\end{gather}
where $\sim$ means in the conjugacy class, $a\notin \mathbb{Z}+\frac{1}{2}$, $m\in \mathbb{C}$ and 
\begin{gather}
     M_{B}=  \left(\begin{array}{cc}
           \frac{\sin \pi(2a-m)}{\sin 2\pi a}e^{-i\nu/2}  & \frac{\sin \pi m}{\sin 2\pi a}  \\ \\
       -\frac{\sin \pi m}{\sin 2\pi a}  &  \frac{\sin \pi(2a+m)}{\sin 2\pi a}e^{i\nu/2}
    \end{array}  \right).
\end{gather}
The parameters $a, \nu$ are called the monodromy data. 

\subsection{Fredholm determinant}\label{subsec:Fred_det}
There are three key steps in writing the tau-function \eqref{eq:IsomTau11} as a Fredholm determinant:
\begin{itemize}
    \item[1.] \hyperref[sec:PantsDecomposition]{pants decomposition},
    \item[2.] \hyperref[sec:CauchyOp]{Cauchy operators},
    \item[3.] \hyperref[sec:DetCauchy]{determinant of the Cauchy operators.}
\end{itemize}
\subsubsection*{1. Pants decomposition} \label{sec:PantsDecomposition}

The  pants decomposition of a one-point torus, i.e cutting the torus along its a-cycle, gives us a 3-point sphere (pair of pants).
\begin{figure}[H]
\begin{center}
\begin{subfigure}{.45\textwidth}
\centering
\begin{tikzpicture}[scale=1.5]
\draw[thick,decoration={markings, mark=at position 0.25 with {\arrow{>}}}, postaction={decorate}] (0,0) circle [x radius=0.5, y radius =0.2];
\node at (0,0) {$M_0$};
\draw(-0.5,0)  to[out=270,in=90] (-1.5,-1.5) to[out=-90,in=-90] (1.5,-1.5);
\draw(0.5,0) to[out=270,in=90] (1.5,-1.5);

\draw(-0.7,-1.4) to[out= -30,in=210] (0.7,-1.4);
\draw(-0.55,-1.469) to[out= 30,in=-210] (0.55,-1.469);

\fill[red!30!white] (-0.2,-2.375) to[out=135,in=225] (-0.2,-1.605)
to (0.2,-1.6) to[out=190,in=170] (0.2,-2.373) --cycle;

\draw[dashed,color=black!60!white](-0.2,-2.375) to[out=135,in=225] (-0.2,-1.605)
to (0.2,-1.6) to[out=190,in=170] (0.2,-2.373) --cycle;

\fill[red](-0.2,-2.375) to[out=0,in=0] (-0.2,-1.605)
to (0.2,-1.6) to[out=10,in=-10] (0.2,-2.373) --cycle;

\draw(-0.2,-2.375) to[out=0,in=0] (-0.2,-1.605)
to (0.2,-1.6) to[out=10,in=-10] (0.2,-2.373) --cycle;
\node at ($(0,-0.7)$) {\Large $\mathscr{T}$};
\node at ($(0,-2.6)$) {\color{red}\Large $\mathscr{A}$};
\node at ($(-0.5,-1.8)$) {$\cC_{in}$};
\node at ($(0.7,-1.8)$) {$\cC_{out}$};
\end{tikzpicture}
\caption{Pants decomposition of $C_{1,1}$}
\label{fig:Torus}
\end{subfigure}
\hfill
\begin{subfigure}{.45\textwidth}
\centering
\begin{tikzpicture}[scale=1.5]
\draw[thick,decoration={markings, mark=at position 0.25 with {\arrow{>}}}, postaction={decorate}](0,0) circle[x radius=0.5, y radius =0.2];
\draw[red,thick,decoration={markings, mark=at position 0.25 with {\arrow{>}}}, postaction={decorate}] (-1,-2) circle[x radius=0.5, y radius =0.2];
\draw[red,thick,decoration={markings, mark=at position 0.25 with {\arrow{<}}}, postaction={decorate}] (1,-2) circle[x radius=0.5, y radius =0.2];

\draw(-0.5,0)  to[out=270,in=90] (-1.5,-2);
\draw(0.5,0) to[out=270,in=90] (1.5,-2);
\draw(-0.5,-2) to[out=90,in=180] (0,-1.5) to[out=0, in=90] (0.5,-2);
\node at ($(0,-0.7)$) {\Large $\mathscr{T}$};

\node at ($(0.6,0.3)$) {$M_{0}$};
\node at ($(-1,-2.5)$) {$M_A$};
\node at ($(-1.8,-2.0)$) {$\cC_{in}$};
\node at ($(1,-2.5)$) {$M_{B}^{-1} M_{A} M_{B}$};
\node at ($(1.8,-2.0)$) {$\cC_{out}$};
\end{tikzpicture}
\caption{Trinion}
\label{fig:Trinion}
\end{subfigure}
\end{center}
\vspace{-0.5cm}
\caption{}
\label{fig:TorusTrinion}
\end{figure}
The pants decomposition induces a homomorphism $\phi$ \cite{Goldman2009arXiv0901.1404G} such that the monodromy around $\mathcal{C}_{in}, \mathcal{C}_{out}$ denoted by $M_{in}$, $M_{out}$ respectively are 
\begin{align}
    \phi(M_{A}) =: M_{in}, && \phi(M_0) = M_0, && \phi (M_{B}^{-1} M_{A}^{-1} M_{B}) =: M_{out},
\end{align}
and the monodromy constraint
\begin{gather}
   M_{A}M_0M_{B}^{-1}M_{A}^{-1}M_{B}=1=(M_{A})M_0(M_{B}^{-1}M_{A}M_{B})^{-1}  := M_{in}M_0 M_{out}.
\end{gather}
\begin{definition}\label{def:YcalY}
The solution $\mathcal{Y}_{CM}(z)$, of the linear problems \eqref{linear_system}, can be redefined on the boundary spaces 
\begin{align}\label{eq:def_Hinout}
\mathcal{H}_{in}:=L^{2}\left( \mathcal{C}_{in} \right), && \mathcal{H}_{out}:=L^{2}\left( \mathcal{C}_{out} \right), 
\end{align} 
such that $Y_{CM}(z)$, has diagonal monodromies $M_{A}, M_{A}^{-1}$ respectively around the boundary circles $\cC_{in}$ and $\cC_{out}$ in Figure \ref{fig:TorusTrinion} as 
 \begin{align}\label{eq:YcalY}
    Y_{CM}(z)\vert_{\cC_{in}}:=\mathcal{Y}_{CM}(z)\vert_{\cC_{in}}\in\cH_{in}, && Y_{CM}(z)\vert_{\cC_{out}}:=M_B^{-1}\mathcal{Y}_{CM}(z)\vert_{\cC_{out}}\in\cH_{out}.
\end{align}
\end{definition}
We now associate the following linear system to the pair of pants fig. \ref{fig:TorusTrinion}b: the $2\times 2$ matrix valued function $\widetilde{Y}(z)$ lives on a cylinder with 3 punctures at $-i\infty, 0, i\infty$ respectively
\begin{align}
    \partial_z\widetilde{Y}(z)= -2\pi i\left(  \widetilde{A}_{in}+\frac{\widetilde{A}_0}{1-e^{2\pi i z}} \right)\widetilde{Y}(z), &&  \widetilde{A}_{in}\sim a\sigma_3, && \widetilde{A}_{0}\sim m\sigma_3, \label{eq:linear_system3pt}
\end{align}
with $A_{in}, A_{0}$ ensuring that the monodromies around $\mathcal{C}_{in}, \mathcal{C}_{out}$ hold true. 
The local solutions of the 3-point sphere
\begin{align} \Yt_{in}(z):= \Yt(z)\vert_{\cC_{in}}, &&
    \widetilde{Y}_{out}(z):=e^{2\pi i  \bm{{\nu}}}\sigma_1 \widetilde{Y}_{in}(-z) \sigma_1, \label{eq:3pt_local_sol}
\end{align} 
are described by the Wronskians of the hypergeometric functions, where
\eq{\label{eq:in_hyp}
\Yt_{in}(z)=(1-e^{-2\pi iz})^{m}\operatorname{diag}(e^{2\pi i az},e^{-2\pi i az})\times\\\times
\begin{pmatrix}
{}_2F_1(m,m-2a,-2a,e^{-2\pi iz})&
-\frac{m}{2a}{}_2F_1(1+m,m-2a,1-2a,e^{-2\pi iz})\\\frac{m e^{-2\pi iz}}{2a+1}{}_2F_1(1+m,1+m+2a,2+2a,e^{-2\pi iz})&{}_2F_1(m,1+m+2a,1+2a,e^{-2\pi iz})
\end{pmatrix}.
}
is the local behavior of the solution to the associated linear system \eqref{eq:linear_system3pt} for $z\rightarrow-i\infty$, normalized so that the monodromy around $-i\infty$ is $e^{2\pi ia\sigma_3}$. The above function has a  well-defined series expansion in $e^{-2\pi iz}$, is convergent for $|e^{-2\pi iz}|<1$, and ${_2F_1}$ are hypergeometric functions,
We will soon see that the tau-function \eqref{eq:IsomTau11} will be completely determined by the functions $\widetilde{Y}_{in}, \widetilde{Y}_{out}$.

\subsubsection*{2. Cauchy operators}\label{sec:CauchyOp}
Let us introduce two Cauchy operators $\cP_{\oplus}$, $\cP_{\Sigma}$ respectively in terms of the solutions of the 3-point sphere $\widetilde{Y}(z)$ \eqref{eq:linear_system3pt} and the linear systems on the one-point torus $Y_{CM}(z,\tau)$ \eqref{eq:YcalY}.
We define $$\mathcal{H} := \mathcal{H}_{in} \oplus \mathcal{H}_{out} = \mathcal{H}_{+} \oplus \mathcal{H}_{-}$$ with $\mathcal{H}_{in}$, $\mathcal{H}_{out}$ defined in \eqref{eq:def_Hinout} and $\mathcal{H}_{\pm}$ defining the positive and negative Fourier modes.
 \begin{itemize}
     \item The  operator $ \cP_{\oplus}: \cH \rightarrow \cH $ is defined as
\begin{gather}
    \left(\cP_{\oplus} f\right)(z) = \int_{\cin \cup \cout} dw\, \frac{\Yt(z) \Yt (w){^{-1}}}{1-e^{-2\pi i(z-w)}} f(w) \nonumber \\
    = \int_{\mathcal{C}} dw\, \frac{\Yt(z) \Yt (w){^{-1}}}{1-e^{-2\pi i(z-w)}} f(w). \label{pp}
\end{gather}
Note that the above is the Cauchy operator in cylindrical coordinates. For $z\sim w$
 \begin{equation}\frac{1}{1- e^{-2\pi i(z-w)}} = \frac{1}{2\pi i (z-w)} + \frac{1}{2} + \frac{2\pi i }{12}(z-w)+\mathcal{O}\left((z-w)^2\right). \label{eq:ExpExpansion}\end{equation}
It is straightforward to verify that $\cP_{\oplus}$ is indeed projective, i.e $ \cP_{\oplus}^2 = \cP_{\oplus}$, and $\textrm{ker}~\cP_{\oplus} = \cH_{-}$. 
\end{itemize}
A function $f(z) \in \mathcal{H}$ has the following splitting in terms of positive and negative Fourier modes respectively
\begin{gather}
    f(z) =  \begin{pmatrix}f_{in,-}\\f_{out,+}  \end{pmatrix} \oplus  \begin{pmatrix}f_{in,+}\\f_{out,-}  \end{pmatrix}.
\end{gather}
 The operator $\cP_{\oplus}$ then has the following explicit form
\begin{gather}
    \label{eq:Pplusabcd}
  (\mc P_\oplus f)(z)=\begin{pmatrix}(\mc P_\oplus f)_{in,-}\\(\mc P_\oplus f)_{out,+}  \end{pmatrix}\oplus\begin{pmatrix}  (\mc P_\oplus f)_{in,+}\\(\mc P_\oplus f)_{out,-}
  \end{pmatrix}
  \nonumber \\=
  \begin{pmatrix}f_{in,-}\\f_{out,+}  \end{pmatrix}
  \oplus\begin{pmatrix}{\sf a}&{\sf b}\\{\sf c}&{\sf d}\end{pmatrix}\begin{pmatrix}f_{in,-}\\f_{out,+}\end{pmatrix},
\end{gather}
where ${\sfa}, {\sfb}, {\sfc}, {\sfd} $ are the components of $\cP_\oplus$ with respect to the decomposition $\cH=\cH_{in} \oplus\cH_{out} $:
{\small \begin{gather}
  ({\sfa} g)(z)=\oint_{\mc C_{in}}dw\frac{\Yt_{in}(z)\Yt_{in}(w)^{-1}-\mathbb{1}}{1-e^{-2\pi i(z-w)}}g(w), \,\, z\in\cin, \label{eq:kernela}\\
  ({\sfb} g)(z)=\oint_{\mc C_{out}}dw\frac{\Yt_{in}(z)\Yt_{out}(w)^{-1}}{1-e^{-2\pi i(z-w)}}g(w), \,\, z\in\cin, \label{eq:kernelb}\\
  ({\sfc} g)(z)=\oint_{\mc C_{in}}dw\frac{\Yt_{out}(z)\Yt_{in}(w)^{-1}}{1- e^{-2\pi i(z-w)}}g(w), \,\, z\in\cout, \label{eq:kernelc}\\
  ({\sfd} g)(z)=\oint_{\mc C_{out}} dw \frac{\Yt_{out}(z)\Yt_{out}(w)^{-1}-\mathbb{1}}{1- e^{-2\pi i(z-w)}}g(w), \,\, z\in\cout.
  \label{eq:kernels}
\end{gather}}
The functions $\Yt_{in},\Yt_{out}$ are the local solutions of the three-point problem \eqref{eq:3pt_local_sol} around $\mp i\infty$.

 \begin{itemize}
     \item  The operator $\cP_{\Sigma}:\cH \rightarrow \cH$ is defined in terms of the redefined solutions of the linear system \eqref{eq:YcalY} as
\begin{gather} 
    \left(\cP_{\Sigma} f\right)(z) = \int_{\cin \cup \cout} \frac{dw}{2\pi i}\, Y_{CM}(z,\tau)\Xi(z,w; \tau) Y_{CM}(w,\tau)^{-1} f(w) \nonumber\\
    \equiv\int_{\mathcal{C}} \frac{dw}{2\pi i} \, Y_{CM}(z,\tau)\Xi(z,w; \tau) Y_{CM}(w,\tau)^{-1} f(w),\label{ps}
\end{gather}
where the {\it twisted} Cauchy kernel
\begin{align}\label{eq:twistedCauchy} 
    \Xi(z,w; \tau)= \left( \begin{array}{cc}
        \frac{\theta_1(z-w+Q- \rho)\theta_1'(0)}{\theta_1(z-w)\theta_1(Q-\rho)} &  0 \\
        0 & -\frac{\theta_1(z-w-Q- \rho)\theta_1'(0)}{\theta_1(z-w)\theta_1(Q+\rho)}
    \end{array} \right),
\end{align}
and 
\begin{gather}
    \Xi(z+\tau,w; \tau)=e^{-2\pi iQ\sigma_3+2\pi i\rho} \Xi(z,w; \tau), \nonumber \\
    \Xi(z,w+\tau; \tau)=\Xi(z,w; \tau) e^{2\pi i Q\sigma_3-2\pi i\rho},\label{eq:TauShiftXi2}
\end{gather}
where $\rho$ is a parameter encoding a $U(1)$ B-cycle monodromy of the twisted Cauchy kernel. 
\begin{remark}
Cauchy kernels on a torus are not unique but are determined by their divisor. In the present case, the Cauchy kernel has one simple pole at $z=w$ and is explicitly dependent on the transcendent $Q(\tau)$. On the contrary, there are other Cauchy kernels that do not depend on the solutions to the integrable equations being studied but have an extra pole. Such kernels have been studied in \cite{bertola2021pade, rodin2013riemann, bikbaev1988asymptotics}. 
\end{remark}
 \end{itemize}
 
 \begin{itemize}
        \item For $z\sim w$,
\begin{equation}\label{Xi_expansion}
    \begin{split}\Xi(z,w; \tau) = \frac{\mathbb{1}}{z-w} &+ \diag\left[ \frac{\theta_{1}'(Q-\rho)}{\theta_{1}(Q-\rho)}, -\frac{\theta_{1}'(Q+\rho)}{\theta_{1}(Q+\rho)} \right]\\
    &+\frac{1}{2}(z-w)\,\diag\left[\frac{\theta_1''(Q-\rho)}{\theta_1(Q-\rho)}, \frac{\theta_1''(Q+\rho)}{\theta_1(Q+\rho)}\right]\\
    &  -\frac{\mathbb{1}}{6}(z-w)\frac{\theta_1'''}{\theta_1'} + \mc{O}\left( (z-w)^2 \right).  \end{split}
    \end{equation}
    
  As before, one can easily verify that $\cP_{\Sigma}^2 = \cP_{\Sigma}$, and the space of functions on the annulus $ \cH_{\mathscr{A}}= \textrm{ker}\cP_{\Sigma}$ (assuming $\cP_{\Sigma}$ is invertible).

   \item Moreover, $
        \cP_{\oplus}\cP_{\Sigma} = \cP_{\Sigma} ,\quad \cP_{\Sigma}\cP_{\oplus}= \cP_{\oplus}$ 
Therefore, the space of functions on the trinion $\mathscr{T}$ in figure \ref{fig:TorusTrinion} is defined as
    \begin{gather}
   \cH_{\mathscr{T}}:= \textrm{im}~ \cP_{\oplus} = \textrm{im}~ \cP_{\Sigma} \label{eq:H_T},
\end{gather}
such that the Hilbert space
\begin{gather}
    \mathcal{H} =  \cH_{\mathscr{T}} \oplus  \cH_{\mathscr{A}}
\end{gather}
with $ \cH_{\mathscr{A}}$ being the space of functions on the annulus $\mathscr{A}$ in figure \ref{fig:TorusTrinion}. 
    \end{itemize}

\subsubsection*{3. Determinant of Cauchy operators} \label{sec:DetCauchy}
As a last step to expressing the tau-function \eqref{eq:IsomTau11} as a Fredholm determinant, let us define the operator $K$ such that
\begin{gather} \label{eq:def_Fred_det}
   \det \left[ \mathbb{1} - K(\tau)\right]:=\det_{\cH_+}\left[\cP_{\Sigma,+}^{-1}\cP_{\oplus,+} \right] = \det \left[ \mathbb{1} - \begin{pmatrix}
   \nabla^{-1}{\sfc}& \nabla^{-1}{\sfd} \nabla\\{\sfa}&{\sfb}\nabla
  \end{pmatrix} \right],\quad \cP_{\cdot,+}:=\cP_\cdot\vert_{\cH_+}, 
\end{gather}
and $\nabla$ is a shift operator defined as 
\begin{gather}
    \nabla g(z) = e^{2\pi i \rho} g(z-\tau).
\end{gather}
With some straightforward analysis (refer to the proof of proposition 1 in \cite{del2020isomonodromic}) one can see that the operator $K(\tau)$ depends solely on the operators \eqref{eq:kernela}-\eqref{eq:kernels} shifted by the action of the operator $\nabla$.
\begin{theorem}
The logarithmic derivative of the Fredholm determinant \eqref{eq:def_Fred_det} gives back the isomonodromic tau-function \eqref{eq:IsomTau11} 
\begin{equation}
\begin{split} 
     2\pi i \partial_{\tau} \log\det\left[\mathbb{1}- K  \right] =  2\pi i &\partial_{\tau} \log \T_{CM} -(2\pi i)^{2} {a}^2 -\frac{(2\pi i)^2}{6}\\ & + 2\pi i \frac{d}{d\tau} \log \left( \frac{\theta_{1}(Q-{\rho}) \theta_{1}(Q+{\rho})}{\eta(\tau)^2}\right).
 \end{split}\label{proof:thm1}
\end{equation}
\end{theorem}
\begin{sproof}
Take the logarithmic derivative of the determinant and express the Cauchy operators in their integral form (refer to Theorem 1 in \cite{del2020isomonodromic} for a detailed proof).
\end{sproof}

One of the merits of the determinant representation is that the transcendent $Q(\tau)$ of the elliptic form of Painlev\'e VI can be explicitly expressed in terms of the determinant
    \begin{gather}\label{eq:det_transcendent}
        \frac{\theta_{3}(2Q(\tau)\vert 2\tau)}{\theta_{2}(2Q(\tau)\vert 2\tau)} = i e^{3 i \pi \tau/2} \frac{\det\left( \mathbb{1}- K\vert_{\rho = \frac{1}{4} + \frac{\tau}{2}} \right)}{\det\left( \mathbb{1}- K\vert_{\rho = \frac{1}{4}} \right)},
    \end{gather}
    which is derived using the identity
    \begin{gather}\label{id:theta1_23}
        \theta_{1}(x-y\vert \tau)\theta_{1}(x+y\vert \tau) = \theta_{3}(2x\vert 2\tau) \theta_{2}(2y\vert 2\tau)-\theta_{2}(2x\vert 2\tau) \theta_{3}(2y\vert 2\tau),
    \end{gather}
    where 
    \begin{align}\label{series:theta23}
        \theta_{2}(z\vert \tau) &= \sum_{n\in \mathbb{Z}} q^{(n+1/2)^2}/2 e^{2\pi i z(n+1/2)},\nonumber \\
        \theta_{3}(z\vert \tau) &= \sum_{n\in \mathbb{Z}} q^{n^2}/2 e^{2\pi i zn}.
    \end{align}
The above equation can be obtained from equations 3.56 and F.3 in \cite{Bonelli:2019boe}). In this sense, the Fredholm determinant is the 'true' tau-function associated to the equation \eqref{eq:EllPainleve}.

\subsection{Minor expansion and combinatorial expression}

The minor expansion of the Fredholm determinant \eqref{eq:def_Fred_det} contains the conformal block structure, and it is useful to make the residue matrix around the puncture $z=0$ rank-1\footnote{The reason behind this is computational and is detailed in \cite{GL, del2020isomonodromic} } through a simple gauge transformation to make the precise connection with the Nekrasov-Okounkov functions (see remark \eqref{remark:partition}). 

The Lax matrix $L_{1}$ in \eqref{linear_system} has a rank-2 residue at the puncture $z=0$:
\begin{equation}
        L_{1}(z)=\left(\begin{array}{cc}
        P & mx(2Q,z) \\
        mx(-2Q,z) & -P
    \end{array}\right)= \frac{m\sigma_1}{z}+\mathcal{O}(1).
\end{equation}
The scalar gauge transformation
\begin{align} \label{eq:11_Transformation}
    L_{1}(z)\rightarrow \widetilde{L}_{1} := L_{1}(z)-\lambda(z)^{-1}\partial_z \lambda(z), && \lambda(z)=\theta_1(z)^m, \end{align}
makes the residue at the puncture rank-1
\begin{equation}\label{eq:gaugetransfLax_11}
        \widetilde{L}_{1}(z)=\left(\begin{array}{cc}
        P-m\frac{\theta_1'(z)}{\theta_1(z)} & mx(2Q,z) \\
        mx(-2Q,z) & -P-m\frac{\theta_1'(z)}{\theta_1(z)}
    \end{array}\right)= \frac{m}{z}\left(\begin{array}{cc}
        -1 & 1 \\
        1 & -1
    \end{array} \right)+ \mathcal{O}(1).
\end{equation}
The above gauge transformation, through the periodicity of theta functions, induces an additional scalar factor $g_B(z)$ to the B-cycle monodromy and the A-cycle monodromy remains unchanged:
\begin{gather}
 \lambda(z+\tau)=\theta_1(z+\tau)^{m}=e^{-2\pi i\left(z+\frac{\tau+1}{2}\right)m}\lambda(z) := g_{B}(z) \lambda(z), \label{eq:LambdaCMB} \\
    \lambda(z+1)=\theta_1(z+1)^{m}=e^{i\pi m}\lambda(z):=g_A\lambda(z).
\end{gather}
The monodromy around $z=0$ also gains a scalar factor 
\begin{equation}\label{eq:LambdaCM1}
    g_1= e^{-2\pi im}
\end{equation} 
due to the $z$-dependence of $g_B(z)$.
All in all, monodromy matrices around $\mathcal{C}_{in}, \mathcal{C}_{out}$ are
\begin{align}
    M_{in} = e^{2\pi i \,\diag\left( a-\frac{m}{2},-a-\frac{m}{2} \right)} && M_{out} = e^{2\pi i \, \diag\left( a+\frac{m}{2},-a+\frac{m}{2} \right)},
\end{align}
and the parameter $\rho$ transforms to 
\begin{align}  \widetilde{\rho} = \rho -\frac{m (\tau+1)}{2}. \label{eq:Lambda01}
\end{align}
We now define the monodromy exponents $\vec{\sigma}_{in}$, $\vec{\sigma}_{out}$ as
\begin{align}
\vec{\sigma}_{in}:=\left(a-\frac{m}{2},-a-\frac{m}{2} \right), && \vec{\sigma}_{out}:=\left(a+\frac{m}{2},-a+\frac{m}{2} \right). \label{eq:sigma12}
\end{align}
The tau-function $\widetilde{\T}_{CM}$ associated to the Lax matrix \eqref{eq:gaugetransfLax_11} is then
\begin{equation}\label{eq:tildeTCM}
    2\pi i\partial_\tau\log\widetilde{\T}_{CM} := \frac{1}{2} \oint_{A}dz \tr \widetilde{L}_{1}^{2}.
\end{equation}    
\begin{proposition}
The tau-function $\T_{CM}$ in \eqref{proof:thm1} is related to the tau-function $\widetilde{\T}_{CM}$ in \eqref{eq:tildeTCM} associated to the rank-1 Lax matrix  \eqref{eq:11_Transformation} as
\begin{equation} \label{eq:prop_tildetau_tau}
\T_{CM}(\tau) =\widetilde{\T}_{CM}(\tau) \left(\eta(\tau)e^{\frac{i\pi\tau}{6}}\right)^{-2m^2},
\end{equation}
where $m$ is the monodromy exponent at the puncture and $\tau$ is the isomonodromic time.
\end{proposition}
\begin{sproof}
The equality \eqref{eq:prop_tildetau_tau} can be proved by taking the logarithmic derivative w.r.t on both sides. Refer to section 4.2 in \cite{del2020isomonodromic} for the complete proof.
\end{sproof}
The construction in the section \ref{subsec:Fred_det} holds for the rank-1 case with the appropriate shifts in monodromy parameters and the parameter $\rho$. 
The relation equivalent to \eqref{proof:thm1} is
\begin{equation}
\begin{split} 
    2\pi i &\partial_{\tau} \log \widetilde{\T}_{CM}  =  2\pi i \partial_{\tau} \log\det\left[\mathbb{1}- \widetilde{K}  \right] +(2\pi i)^{2} \left({a}^2 +\frac{m^2}{4} \right) +\frac{(2\pi i)^2}{6}\\ & + 2\pi i \frac{d}{d\tau} \log \left( \frac{ \eta(\tau)^2 }{\theta_{1}\left(Q-{\rho}+\frac{m(\tau +1)}{2}\right)) \theta_{1}\left(Q+{\rho}- \frac{m(\tau +1)}{2}\right)}\right).
 \end{split}\label{tilde:thm1}
\end{equation}
Now we perform the minor expansion for the rank-1 analogue of the Fredholm determinant $\det\left[\mathbb{1}- \widetilde{K}  \right]$. 

\vspace*{0.2cm}

There are two main steps to write the minor expansion as Nekrasov-Okounkov functions
\begin{itemize}
    \item[1.] \hyperref[sec:basis]{Basis expansion}
    \item[2.] \hyperref[sec:combinatorics]{Combinatorial expression}
\end{itemize}
\subsubsection*{Basis expansion}\label{sec:basis}
The first step in the minor expansion is noting that kernels of the operators $\sfa, \sfb, \sfc, \sfd$ in \eqref{eq:kernels} are
\eqs{\label{eq:abcdYt}
    \sfa (z,w) = \frac{\mathbb{1}-\Yt_{in}(z) \Yt_{in}(w)^{-1}}{1-e^{-2\pi i(z-w)}}, && z,w\in \cin, \\
    \sfb(z,w)\, = \frac{\Yt_{in}(z) \Yt_{out}(w)^{-1}}{1-e^{-2\pi i(z-w)}}, && z\in \cin, \, w\in \cout \\
    \sfc(z,w) = -\frac{\Yt_{out}(z) \Yt_{in}(w)^{-1}}{1-e^{-2\pi i(z-w)}}, && z\in \cout, \, w\in \cin \\
    \sfd(z,w) = \frac{\Yt_{out}(z) \Yt_{out}(w)^{-1}-\mathbb{1} }{1-e^{-2\pi i(z-w)}}, && z,w\in \cout.}
The solution to the three-point problem $\Yt$ is multivalued\footnote{For the ease of notation we the keep the notation of the kernels $\sfa, \sfb, \sfc, \sfd$.} on $\cC_{in},\cC_{out}$, with monodromy determined by $ M_{in},M_{out}$ defined in \eqref{eq:sigma12}, consequently the basis is monodromy dependent, and the  matrix elements of the kernels above are
\begin{equation}
    \begin{split}
\sfa_{\alpha, \beta}(z,w) = \sum_{-r,s\in\mathbb{Z}'_+} \sfa^{-r;\alpha}_{s;\beta} e^{2\pi iz\left(\frac{1}{2}-r+{ \sigma}_{in}^{(\alpha)}\right)} e^{2\pi iw\left(-\frac{1}{2}-s- { \sigma}_{in}^{(\beta)}\right)}, \\
    \sfb_{\alpha, \beta}(z,w) = \sum_{r,s\in\mathbb{Z}'_+} \sfb^{-r;\alpha}_{-s;\beta} e^{2\pi iz\left(\frac{1}{2}-r+{ \sigma}_{in}^{(\alpha)}\right)} e^{2\pi i w\left(-\frac{1}{2}+s- { \sigma}_{out}^{(\beta)}\right)}, \\
    \sfc_{\alpha, \beta}(z,w) = \sum_{r,s\in\mathbb{Z}'_+} \sfc^{r;\alpha}_{s;\beta} e^{2\pi iz\left(\frac{1}{2}+r+{ \sigma}_{out}^{(\alpha)}\right)} e^{2\pi iw\left(-\frac{1}{2}-s- { \sigma}_{in}^{(\beta)}\right)}, \\
    \sfd_{\alpha, \beta}(z,w) = \sum_{r,s\in\mathbb{Z}'_+} \sfd^{r;\alpha}_{-s;\beta} e^{2\pi iz\left(\frac{1}{2}+r+{ \sigma}_{out}^{(\alpha)}\right)} e^{2\pi iw\left(-\frac{1}{2}+s- { \sigma}_{out}^{(\beta)}\right)},\end{split}\label{eq:ker_abcd}\end{equation}
with $\alpha,\beta=1,2$, the Fourier indices are denoted by $r, s$, and $\mathbb{Z}'_+$ is the set of positive half-integers \footnote{The explicit form of the Fourier coefficients $\sfa^{-r;\alpha}_{s;\beta},\sfb^{-r;\alpha}_{-s;\beta},\sfc^{r;\alpha}_{s;\beta},\sfd^{r;\alpha}_{-s;\beta}$
are not relevant for the purposes of this note but are computed in \cite{GL}. }. A submatrix of $\sfa,\sfb,\sfc,\sfd $, of size $i\times j$, is denoted by two unordered sets $\{(r,\alpha)_1,\dots,(r,\alpha)_i\} \in 2^{\mathbb{Z}'_{+} \times \lbrace 1,...,N \rbrace}$ and $\{(s,\beta)_1,\dots,(s,\beta)_j\} \in  2^{\mathbb{Z}'_{+} \times \lbrace 1,2 \rbrace}$. The minors can then be denoted by the multi indices of positive (negative) Fourier modes $I$ ($J$). The minor expansion of the determinant is then indexed by $I,J$ as
\begin{gather}
    \det\left[ \mathbb{1}- \widetilde{K} \right] = \sum_{\left(I,J \right)} \det\left( \begin{array}{cc}\left( \sfa \right)^{I}_{J} &\left( \sfb \nabla \right)^{I}_{I} \\  \\
    \left( \nabla^{-1} \sfc \right)^{J}_{J} & \left(\nabla^{-1} \sfd \nabla \right)^{J}_{I}\end{array}  \right). \label{eq:Det_part}
\end{gather}

\subsubsection*{Combinatorial expression}\label{sec:combinatorics}
The multi-indices $(I,J)$ can be viewed as the positions $\sf h(m^{(\alpha)})$ and $\sf p(m^{(\alpha)})$ of 'holes' and 'particles'  respectively, of a coloured Maya diagram $\sf m^{(\alpha)}$, where $\alpha=1,2$. Each particle (hole) carries a positive (negative) unit charge, so that the total charge associated to every Maya diagram is 
\begin{equation}
    \sf Q({\sf m}^{(\alpha)}) := \vert p(m^{(\alpha)}) \vert - \vert h(m^{(\alpha)})\vert.
\end{equation}
Using the notation
\begin{align}\label{eq:charge_maya}
    \vec{\sf m}:=\left(\sf{m}^{(1)},\sf{m}^{(2)} \right), && \vec{\sf Q}:=\left(\sf{Q}^{(1)},\sf{Q}^{(2)} \right),
\end{align}
the total charge is
\eq{\label{eq:totalchargeMaya}
    \sfQ:=\sf{Q}^{(1)}+\sf{Q}^{(2).}
    }
Furthermore, there is a one-one correspondence between Maya diagrams and charged Young diagrams $\left(\sf Y,\sf Q\right)$. Consequently, the minors can be labeled by $2$-tuples of charged partitions $\left(\vec{\sfY},\vec{\sfQ}\right)$. 
Explicitly, the minor
\begin{gather}
     \det\left( \begin{array}{cc}\left( \sfa \right)^{I}_{J} &\left( \sfb \nabla \right)^{I}_{I} \\  \\
    \left( \nabla^{-1} \sfc \right)^{J}_{J} & \left(\nabla^{-1} \sfd \nabla \right)^{J}_{I}\end{array} \right) \nonumber \\
    = \det\left( \begin{array}{cc}\left( \sfa \right)^{I}_{J} &\left( \sfb  \right)^{I}_{I} \\  \\
    \left(  \sfc \right)^{J}_{J} & \left( \sfd \right)^{J}_{I}\end{array} \right) ~\textrm{exp}\left\{\sum_{(r,\alpha)\in J} \left[-2\pi i \left(\rho-\frac{\tau}{2}- m \tau  \right) +2\pi i \tau\left(r + \sigma_{1}^{(\alpha)} \right)\right]\right\} \nonumber\\
    \times\exp\left\{\sum_{(s,\beta)\in I} \left[2\pi i \left(\rho-\frac{\tau}{2}-m \tau\right) + 2\pi i \tau\left(s - \sigma_{1}^{(\beta)}\right) \right] \right\} \nonumber \\
    = \det\left( \begin{array}{cc}\left( \sfa \right)^{I}_{J} &\left( \sfb  \right)^{I}_{I} \\  \\
    \left( \sfc \right)^{J}_{J} & \left( \sfd \right)^{J}_{I}\end{array} \right) e^{2\pi i\tau\left[\frac{1}{2} \left({\vec {\sf Q} }+ {\vec \sigma}_{1}  \right)^2 - \frac{1}{2} {\vec \sigma}_{1}^2 + \vert \vec{\sf Y}\vert\right]  - 2\pi i \left(\rho - \frac{\tau}{2}- m \tau  \right) {\sf Q} }.\label{eq:Zn_block}
\end{gather}
Here are few more details about the above computation.
\begin{itemize}
    \item[1.] To obtain the second line of \eqref{eq:Zn_block}, note that $\sigma_1$ is the monodromy exponent on $\cC_{in}$, and the monodromy exponent on $\cC_{out}$ is $\sigma_1+ m $.
    \item[2.] To obtain the last line in \eqref{eq:Zn_block}, we use the following equalities:
\begin{gather}
\label{eq:srQY}
\sum_{l} r_{l} + \sum_{k} s_{k} = \frac{\sf Q(m)^2}{2} + \vert \vec{\sf Y} \vert , \qquad \#r-\#s= \sf Q(m), 
\end{gather}
which can be read off from Figure \ref{fig:maya} noting that the $r$'s and $s$'s are to the left and right sides of the axis respectively.
\begin{figure}[H]
\centering
\includegraphics[width=10cm]{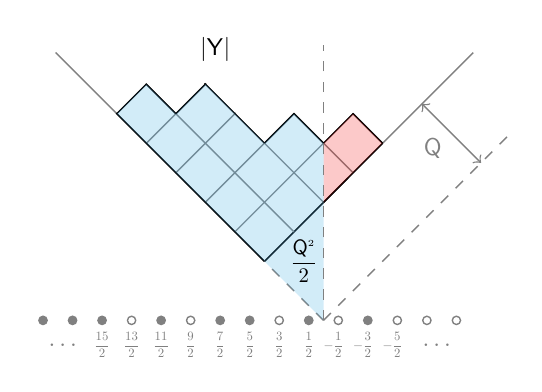}
\caption{Pictorial proof of \eqref{eq:srQY} \label{fig:maya}}
\end{figure}
As an example\footnote{Since $s\in I$, the hole positions in the corresponding Maya diagram $\sf{m}$ are  ${\sf h(m)}= \left\{ -s_{1},...,-s_{k} \right\}$, and since $r\in J$, the particle positions are ${\sf p(m)} =\left\{ r_1,...,r_l  \right\} $. } , in the Figure \ref{fig:maya}, ${\sf p(m)}=\left\{ \frac{13}{2}, \frac{9}{2}, \frac{3}{2} \right\}$, ${\sf h(m)}=\left\{ \frac{-3}{2} \right\}$. $\vert Y \vert$ is the $\#$boxes in the Young diagram which in the present example is 12. The charge $\sf Q(m)=2$, $\sum r$ is the blue area and $\sum s$ is the red area in the Figure \ref{fig:maya}. 
\end{itemize}

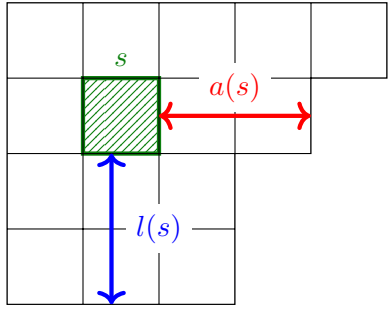
\begin{figure}[H]
    \centering
\begin{tikzpicture}[scale=2.5]
\draw[ultra thick,darkspringgreen,pattern color=darkspringgreen,pattern=north east lines] (0.4,-0.4) rectangle (0.8,-0.8) (0.6,-0.3) node{$s$};
\draw[clip](0,0) -- ++(2,0) -- ++(0,-0.4)-- ++(-0.4,0)-- ++(0,-0.4)-- ++(-0.4,0)-- ++(0,-0.8)--(0,-1.6) --cycle;
\draw[scale=0.4](0,-1.6*3) grid (5,5);
\draw[red](1.2,-0.45) node[fill=white](){ $a(s)$} ;
\draw[ultra thick,red,<->] (0.8,-0.6)--(1.6,-0.6) ;
\draw[blue] (0.8,-1.2) node[fill=white](){ $l(s)$};
\draw[ultra thick,blue,<->] (0.55,-0.8) to (0.55,-1.6)  ;
\end{tikzpicture}
    \caption{Arm and leg length}
    \label{fig:armleg}
\end{figure}

\begin{remark}\label{remark:partition}
\hspace{5cm}
\begin{itemize}
    \item[1.] The determinant in \eqref{eq:Zn_block} is often called the trinion partition function $Z^{{\sfY},{\sfQ}}_{{\sfY},{\sfQ}}\left( \mathscr{T} \right)$:
\begin{gather}
    Z^{{\sfY},{\sfQ}}_{{\sfY},{\sfQ}}\left( \mathscr{T} \right) =Z^{I,J}_{I,J} \left(\mathscr{T} \right):=  \det\left( \begin{array}{cc}\left( \sfa \right)^{I}_{J} &\left( \sfb \right)^{I}_{I} \\  \\
    \left( \sfc \right)^{J}_{J} & \left(\sfd  \right)^{J}_{I}\end{array}  \right) \label{eq:YQ_Z}.
\end{gather}
\item[2.] The expressions such as \eqref{eq:Zn_block} are valid for linear systems of any rank. However, the trinion partition functions \eqref{eq:YQ_Z} are explicitly known in terms of Nekrasov-Okounkov functions only in the case where the Lax matrix residues are of rank-1 \cite{GL,Gavrylenko:2018fsm}. 
\end{itemize}

\end{remark}
Using the expressions for $Z^{\sfY,\sfQ }_{\sfY,\sfQ }(\mathscr{T })$ computed in \cite{GL,Gavrylenko:2018fsm} for the rank-1 case, the determinant \eqref{eq:YQ_Z} is
\begin{gather}
  \det\left( \begin{array}{cc}\left( \sfa \right)^{I}_{J} &\left( \sfb \right)^{I}_{I} \\  \\
    \left( \sfc \right)^{J}_{J} & \left( \sfd  \right)^{J}_{I}\end{array} \right) = e^{2\pi i\vec{\sfQ}\cdot\vec{\nu}} 
\frac{Z_{pert}\left(\vec{a}+\vec{\sf Q},\vec{a}+\vec{\sf Q}+m \right)}{Z_{pert}\left(\vec{a},\vec{a}+m\right)}Z_{inst}\left(\vec{a}+\vec{\sf Q},\vec{a}+\vec{\sf Q}+m\vert\vec{\sf Y},\vec{\sf Y} \right), \label{eq:Z_partition}
    \end{gather}
where $\vec{a} = (a,-a)$ and the functions
\begin{equation}\label{eq:Zpertk}
    Z_{pert}\left(\sigma,{\mu}\right):= \prod_{\alpha,\beta=1}^N\frac{ G\left(1+ \sigma^{(\alpha)}- \mu^{(\beta)}\right)}{G\left(1+\sigma^{(\alpha)}-\sigma^{(\beta)}\right)},
\end{equation}
$G(x)$ being the Barnes' G-function, and
{\begin{gather}
    Z_{inst}\left({\sigma},{\mu}\vert\vec{\sfY},\vec{\sf{W}}\right)
    :=\prod_{\alpha,\beta=1}^N\frac{Z_{\textrm{\sf bif}} \left( \sigma^{(\alpha)}- \mu^{(\beta)} \vert \sfY^{(\alpha)}, \sf{W}^{(\beta)} \right)}{Z_{\textrm{\sf bif}} \left( \sigma^{(\alpha)}- \sigma^{(\beta)}  \vert \sfY^{(\alpha)}, \sfY^{(\beta)} \right)}, \label{eq:Zinstdef}
\end{gather} }
with
\begin{equation}\label{eq:Zbif}
    Z_{\sf bif}\left(x|\sfY',\sfY \right):=\prod_{\Box\in\sfY}\left(x+1+a_{\sfY'}(\Box)+l_{\sfY}(\Box) \right)\prod_{\Box'\in\sfY'} \left(x-1-a_{\sfY}(\Box')-l_{\sfY'}(\Box') \right).
\end{equation}
Note that, $\sigma,\mu\in\mathbb{C}^2$, $\vec{\sfY},\vec{\sf{W}}\in\mathbb{Y}^2$, and $a_{\sfY}(\Box)$ and $l_{\sfY}(\Box)$ denote respectively the arm and leg length of the box $\Box$ in the Young diagram $\sfY$, as in figure \ref{fig:armleg}. 

\begin{theorem} The combinatorial expression of the tau-function $\T_{CM}$ reads
\begin{equation}
\begin{split}
    \T_{CM} 
    & = \frac{\left(\eta(\tau)e^{-\frac{i\pi\tau}{12}}\right)^{2-2m^2} e^{-2\pi i\left[\rho-\frac{\tau}{2}\left(m+\frac{1}{2}\right)- \frac{m}{2}\right]}}{\theta_1\left(Q+\rho-\frac{m(\tau+1)}{2}\right)\theta_1\left(Q-\rho+\frac{m(\tau+1)}{2}\right)}
\sum_{\vec{\sfQ}}\sum_{\vec{\sfY}\in\mathbb{Y}^2}e^{2\pi i\tau\left[\frac{1}{2}(\vec{\sfQ}+\vec{a})^2+|\vec{\sfY}|\right]}\\
&\times e^{2\pi i\left[\vec{\sfQ}\cdot\vec{\nu}-\sfQ\left(\rho-\frac{m(\tau+1)}{2}-\frac{\tau}{2}\right)\right]} 
\frac{Z_{pert}\left(\vec{a}+\vec{\sf Q},\vec{a}+\vec{\sf Q}+m \right)}{Z_{pert}\left(\vec{a},\vec{a}+m\right)}Z_{inst}\left(\vec{a}+\vec{\sf Q},\vec{a}+\vec{\sf Q}+m\vert\vec{\sf Y},\vec{\sf Y} \right). \\
\end{split} \label{Thm:comb}
\end{equation}
where $a,\nu$ are monodromy data, $t$ is the isomonodromic time, $m$ is the parameter from the equation, $Z_{pert}, Z_{inst}$ are combinatorial objects defined in \eqref{eq:Zpertk}, \eqref{eq:Zinstdef}.
\end{theorem}

\begin{proof}
Combine \eqref{eq:Z_partition}, \eqref{eq:Zn_block}, \eqref{eq:Det_part}, \eqref{tilde:thm1}, \eqref{eq:prop_tildetau_tau} and integrate on both sides\footnote{Refer to eq 2.45 in \cite{del2020isomonodromic} for the full computation.}. 
\end{proof}

\subsection{Relation to conformal blocks}
Due to the AGT correspondence \cite{onAGT}, the expression \eqref{Thm:comb}, in terms of Nekrasov-Okounkov functions can be put in the form \eqref{tau:fourier} as a discrete Fourier transform of conformal blocks with a few manipulations.
\begin{proposition} The tau-function can be expressed as a discrete Fourier transform of the conformal blocks as 
\begin{align}
\mathcal{T}_{CM} & = \frac{\eta(\tau)^{2}}{\theta_1\left(Q-\rho\right)\theta_1\left(Q+\rho\right)}
\sum_{n,k \in \mathbb{Z}}e^{2\pi i\nu n} q^{\left(k + n/2\right)^2}e^{4\pi i(k+n/2) \left(\rho+1/2\right)} \mathcal{B}(a+n/2,m,q),
\end{align}
where, $a,\nu$ are the monodromy data, $Q$ is the transcendent, $\tau $ is the isomonodromic time, $q=e^{2\pi i \tau}$, $m$ is a parameter of the equation \eqref{eq:EllPainleve}.

\end{proposition}

\begin{proof}
Starting with the expression of the tau-function in terms of the Nekrasov-Okounkov functions \eqref{Thm:comb}
\begin{equation}
\begin{split}
    \T_{CM} 
    & = \frac{\left(\eta(\tau)e^{-\frac{i\pi\tau}{12}}\right)^{2-2m^2} e^{-2\pi i\left[\rho-\frac{\tau}{2}\left(m+\frac{1}{2}\right)- \frac{m}{2}\right]}}{\theta_1\left(Q+\rho-\frac{m(\tau+1)}{2}\right)\theta_1\left(Q-\rho+\frac{m(\tau+1)}{2}\right)}
\sum_{\vec{\sfQ}}\sum_{\vec{\sfY}\in\mathbb{Y}^2}e^{2\pi i\tau\left[\frac{1}{2}(\vec{\sfQ}+\vec{a})^2+|\vec{\sfY}|\right]}\\
&\times e^{2\pi i\left[\vec{\sfQ}\cdot\vec{\nu}-\sfQ\left(\rho-\frac{m(\tau+1)}{2}-\frac{\tau}{2}\right)\right]} 
\frac{Z_{pert}\left(\vec{a}+\vec{\sf Q},\vec{a}+\vec{\sf Q}+m \right)}{Z_{pert}\left(\vec{a},\vec{a}+m\right)}Z_{inst}\left(\vec{a}+\vec{\sf Q},\vec{a}+\vec{\sf Q}+m\vert\vec{\sf Y},\vec{\sf Y} \right). \\
\end{split} \label{Nek_eq1}
\end{equation}
To match with the notation in \cite{Bonelli:2019boe}, we redefine ${\sfQ}^{(1)} = n_{1}, {\sfQ}^{(2)} = n_{2}$ in \eqref{eq:charge_maya}, i.e
\begin{equation}
\vec{\sf Q} =\left(\begin{array}{c} n_{1} \\ n_{2}\end{array}  \right), \quad n_{1}+n_{2} =k,\quad n_{1}-n_{2} =n.
\end{equation}
The terms in the exponents in the above expression \eqref{Nek_eq1}
\begin{align}
\left( \vec{\sf Q} + \vec{a} \right)^2 &= \left( \begin{array}{cc} n_{1}+a  & n_{2}-a\end{array} \right)\left( \begin{array}{c} n_{1}+a  \\ n_{2}-a\end{array} \right) \nonumber\\
&= (n_{1}+a)^2 + (n_{2}-a)^2 \nonumber \\
&= n_{1}^2+n_{2}^2 + 2a^2 + 2a(n_{1}-n_{2}) \nonumber \\
& = \frac{1}{2}\left[ (n_{1}+n_{2})^2 + (n_{1}-n_{2})^2 \right] + 2\left( a+ \frac{(n_{1}-n_{2})}{2} \right)^2 - \frac{(n_{1}-n_{2})^2}{2} \nonumber \\
& = \frac{k^2}{2} + 2\left( a+ \frac{n}{2}  \right)^2, \label{Qpa}
\end{align}
and
\begin{align}
\vec{\sfQ}\cdot\vec{\nu} = \left( \begin{array}{cc} n_{1}  & n_{2}\end{array} \right)\left( \begin{array}{c} \nu \\ -\nu \end{array} \right) = \nu(n_{1}- n_{2}) = \nu n. \label{Qn}
\end{align}
Plugging in \eqref{Qpa}, \eqref{Qn} in \eqref{Nek_eq1}
\begin{equation}
\begin{split}
    \T_{CM} 
    & = \frac{\left(\eta(\tau)e^{-\frac{i\pi\tau}{12}}\right)^{2-2m^2} e^{-2\pi i\left[\rho-\frac{\tau}{2}\left(m+\frac{1}{2}\right)- \frac{m}{2}\right]}}{\theta_1\left(Q+\rho-\frac{m(\tau+1)}{2}\right)\theta_1\left(Q-\rho+\frac{m(\tau+1)}{2}\right)}
\sum_{\vec{\sfQ}}\sum_{\vec{\sfY}\in\mathbb{Y}^2}e^{2\pi i\tau\left[\frac{1}{2}(\vec{\sfQ}+\vec{a})^2+|\vec{\sfY}|\right]}\\
&\times e^{2\pi i\left[\vec{\sfQ}\cdot\vec{\nu}-\sfQ\left(\rho-\frac{m(\tau+1)}{2}-\frac{\tau}{2}\right)\right]} 
\frac{Z_{pert}\left(\vec{a}+\vec{\sf Q},\vec{a}+\vec{\sf Q}+m \right)}{Z_{pert}\left(\vec{a},\vec{a}+m\right)}Z_{inst}\left(\vec{a}+\vec{\sf Q},\vec{a}+\vec{\sf Q}+m\vert\vec{\sf Y},\vec{\sf Y} \right) \nonumber \\
& = \frac{\left(\eta(\tau)e^{-\frac{i\pi\tau}{12}}\right)^{2-2m^2} e^{-2\pi i\left[\rho-m \frac{\left(\tau+1\right)}{2}- \frac{\tau}{4}\right]}}{\theta_1\left(Q+\rho-\frac{m(\tau+1)}{2}\right)\theta_1\left(Q-\rho+\frac{m(\tau+1)}{2}\right)}
\sum_{n,k\in \mathbb{Z}}e^{2\pi i\tau\left[\frac{k^2}{4} + \left( a+ \frac{n}{2}  \right)^2 \right]} e^{2\pi i\left[\nu n- k \left(\rho-\frac{m(\tau+1)}{2}-\frac{\tau}{2}\right)\right]}\\
&\times  \sum_{\vec{\sfY}\in\mathbb{Y}^2} e^{2\pi i\tau |\vec{\sfY}|}
\frac{Z_{pert}\left(\vec{a}+\vec{\sf Q},\vec{a}+\vec{\sf Q}+m \right)}{Z_{pert}\left(\vec{a},\vec{a}+m\right)}Z_{inst}\left(\vec{a}+\vec{\sf Q},\vec{a}+\vec{\sf Q}+m\vert\vec{\sf Y},\vec{\sf Y} \right) .
\end{split}
\end{equation}
Making the shift
\begin{gather}
\rho - \frac{m(\tau+1)}{2} \rightarrow - \rho - 1/2 - \tau/2,
\end{gather}
the above expression reads
\begin{align}
\mathcal{T}_{CM} & = \frac{\left(\eta(\tau)e^{-\frac{i\pi\tau}{12}}\right)^{2-2m^2} e^{-2\pi i\left[-\rho- \frac{1}{2}-\tau/2- \frac{\tau}{4}\right]}}{\theta_1\left(Q-\rho-\frac{1}{2}-\tau/2\right)\theta_1\left(Q+\rho+\frac{1}{2}+\tau/2\right)}
\sum_{n,k\in \mathbb{Z}}e^{2\pi i\tau\left[\frac{k^2}{4} + \left( a+ \frac{n}{2}  \right)^2 \right]} e^{2\pi i\left[\nu n- k \left(-\rho-\frac{1}{2}-\tau/2-\frac{\tau}{2}\right)\right]} \nonumber \\
&\times  \sum_{\vec{\sfY}\in\mathbb{Y}^2} e^{2\pi i\tau |\vec{\sfY}|}
\frac{Z_{pert}\left(\vec{a}+\vec{\sf Q},\vec{a}+\vec{\sf Q}+m \right)}{Z_{pert}\left(\vec{a},\vec{a}+m\right)}Z_{inst}\left(\vec{a}+\vec{\sf Q},\vec{a}+\vec{\sf Q}+m\vert\vec{\sf Y},\vec{\sf Y} \right). \label{eq:tau_CB2}
\end{align}
The following relation holds true for theta functions \cite{NIST_DLMF}
\begin{gather}\label{id:theta270}
\theta_{1}\left( z+ (m+n\tau) \right) = (-1)^{m+n} e^{- i \pi \tau n^2} e^{-2\pi inz} \theta_{1}(z),
\end{gather}
implying that for $m = n = -1/2$:
\begin{gather}
\theta_{1}\left( Q- \rho -1/2-\tau/2 \right) = (-1)^{-1} e^{- i \pi \tau/4} e^{i\pi (Q-\rho)} \theta_{1}(Q-\rho), \label{theta_minus}
\end{gather}
and for $m = n = 1/2$:
\begin{gather}
\theta_{1}\left( Q+ \rho + 1/2+\tau/2 \right) = (-1) e^{- i \pi \tau/4} e^{-i\pi (Q+\rho)} \theta_{1}(Q+\rho). \label{theta_plus}
\end{gather}
Using the above expressions \eqref{theta_minus}, \eqref{theta_plus}, the expression \eqref{eq:tau_CB2} simplifies as
\begin{align}
\mathcal{T}_{CM} & = \frac{\left(\eta(\tau)e^{-\frac{i\pi\tau}{12}}\right)^{2-2m^2} e^{2\pi i\left[\rho+ \frac{1}{2}+\frac{3 \tau}{4}\right]}e^{i \pi \tau/2} e^{2i\pi \rho}}{\theta_1\left(Q-\rho\right)\theta_1\left(Q+\rho\right)}
\sum_{n,k\in \mathbb{Z}}e^{2\pi i\tau\left[\frac{k^2}{4} + \left( a+ \frac{n}{2}  \right)^2 \right]} e^{2\pi i\left[\nu n+k \left(\rho+\frac{1}{2}+\tau\right)\right]}\nonumber \\
&\times  \sum_{\vec{\sfY}\in\mathbb{Y}^2} e^{2\pi i\tau |\vec{\sfY}|}
\frac{Z_{pert}\left(\vec{a}+\vec{\sf Q},\vec{a}+\vec{\sf Q}+m \right)}{Z_{pert}\left(\vec{a},\vec{a}+m\right)}Z_{inst}\left(\vec{a}+\vec{\sf Q},\vec{a}+\vec{\sf Q}+m\vert\vec{\sf Y},\vec{\sf Y} \right) \nonumber \\
& = \frac{\left(\eta(\tau)e^{-\frac{i\pi\tau}{12}}\right)^{2-2m^2} }{\theta_1\left(Q-\rho\right)\theta_1\left(Q+\rho\right)}
\sum_{n,k\in \mathbb{Z}} e^{2\pi i n \nu} e^{2\pi i\tau\left[\frac{k^2}{4}+ k + 1 \right]} e^{2\pi i(k+2) \left(\rho+\frac{1}{2}\right)} e^{2\pi i \tau \left( a+ \frac{n}{2}  \right)^2 }\nonumber \\
&\times  \sum_{\vec{\sfY}\in\mathbb{Y}^2} e^{2\pi i\tau |\vec{\sfY}|}
\frac{Z_{pert}\left(\vec{a}+\vec{\sf Q},\vec{a}+\vec{\sf Q}+m \right)}{Z_{pert}\left(\vec{a},\vec{a}+m\right)}Z_{inst}\left(\vec{a}+\vec{\sf Q},\vec{a}+\vec{\sf Q}+m\vert\vec{\sf Y},\vec{\sf Y} \right). \label{eq:tau_CB3}
\end{align}
Now with the following change of index
\begin{gather}
k \rightarrow 2k+n -2,
\end{gather}
the expression
\begin{align}
\frac{k^2}{4} +k +1 & \rightarrow  \frac{1}{4} (2k+n-2)^2 + 2k+n -2 +1 \nonumber \\
& = k^2+\frac{n^2}{4}+1 + kn-2k-n + 2k+n -1 = \left(k+n/2 \right)^2.
\end{align}
With the above change in the index, the expression \eqref{eq:tau_CB3} reads
\begin{align}
\mathcal{T}_{CM} & = \frac{\left(\eta(\tau)e^{-\frac{i\pi\tau}{12}}\right)^{2-2m^2} }{\theta_1\left(Q-\rho\right)\theta_1\left(Q+\rho\right)}
\sum_{n,k\in \mathbb{Z}}e^{2\pi i\nu n} e^{2\pi i\tau\left(k + n/2\right)^2}e^{4\pi i(k+n/2) \left(\rho+1/2\right)} e^{2\pi i \tau\left( a+ \frac{n}{2}  \right)^2 } \nonumber\\
&\times  \sum_{\vec{\sfY}\in\mathbb{Y}^2} e^{2\pi i\tau |\vec{\sfY}|}
\frac{Z_{pert}\left(\vec{a}+\vec{\sf Q},\vec{a}+\vec{\sf Q}+m \right)}{Z_{pert}\left(\vec{a},\vec{a}+m\right)}Z_{inst}\left(\vec{a}+\vec{\sf Q},\vec{a}+\vec{\sf Q}+m\vert\vec{\sf Y},\vec{\sf Y} \right).
\end{align}
The toric conformal block is defined as 
\begin{align}
\tr_{\mathcal{V}_{a+n/2}} \left( q^{L_{0}} V_{m}(0) \right) & = \mathcal{B}(a+n/2,m,q) := \left(\eta(\tau)e^{-\frac{i\pi\tau}{12}}\right)^{-2m^2} e^{-\frac{i\pi\tau}{6}}
 e^{2\pi i \tau\left( a+ \frac{n}{2}  \right)^2 } \nonumber\\
&\times  \sum_{\vec{\sfY}\in\mathbb{Y}^2} e^{2\pi i\tau |\vec{\sfY}|}
\frac{Z_{pert}\left(\vec{a}+\vec{\sf Q},\vec{a}+\vec{\sf Q}+m \right)}{Z_{pert}\left(\vec{a},\vec{a}+m\right)}Z_{inst}\left(\vec{a}+\vec{\sf Q},\vec{a}+\vec{\sf Q}+m\vert\vec{\sf Y},\vec{\sf Y} \right). 
\end{align}
Finally in terms of the conformal blocks defined above, 
\begin{align}\label{proofeq:cb}
\mathcal{T}_{CM} & = \frac{\eta(\tau)^{2}}{\theta_1\left(Q-\rho\right)\theta_1\left(Q+\rho\right)}
\sum_{n,k\in \mathbb{Z}}e^{2\pi i\nu n} q^{\left(k + n/2\right)^2}e^{4\pi i(k+n/2) \left(\rho+1/2\right)} \mathcal{B}(a+n/2,m,q).
\end{align}
\end{proof}
The relation \eqref{proofeq:cb} can be simplified further. The series
\begin{align}
    \sum_{n,k\in \mathbb{Z}}e^{2\pi i\nu n} q^{\left(k + n/2\right)^2}e^{4\pi i(k+n/2) \left(\rho+1/2\right)} \mathcal{B}(a+n/2,m,q)
\end{align}
can be split into the sum over the odd and even values of $n$:
\begin{align}
    & \sum_{k\in \mathbb{Z}, n \in 2\mathbb{Z}+1}e^{4\pi i\nu (n+1/2)} q^{\left(k + n + 1/2\right)^2}e^{4\pi i(k+n+1/2) \left(\rho+1/2\right)} \mathcal{B}(a+n+1/2,m,q)\nonumber \\
    &+  \sum_{k\in \mathbb{Z}, n' \in 2 \mathbb{Z}}e^{4\pi i\nu n'} q^{\left(k + n'\right)^2}e^{4\pi i(k+n') \left(\rho+1/2\right)} \mathcal{B}(a+n',m,q) \nonumber \\
    =& \sum_{n \in 2\mathbb{Z}+1}e^{4\pi i\nu (n+1/2)}  \mathcal{B}(a+n+1/2,m,q)\sum_{l=k+n\in \mathbb{Z}} q^{\left(l+ 1/2\right)^2}e^{4\pi i(l+1/2) \left(\rho+1/2\right)}\nonumber \\
    &+  \sum_{n' \in 2 \mathbb{Z}}e^{4\pi i\nu n'}\mathcal{B}(a+n',m,q) \sum_{l'=k+n'\in \mathbb{Z}} q^{l'^2}e^{4\pi i l' \left(\rho+1/2\right)} \nonumber \\
    &\mathop{=:}^{\eqref{series:theta23}}  \Bigg(
Z^{D}_{1/2}(\tau)\theta_{3}\left(2\rho\vert 2\tau\right)  -Z^{D}_{0}(\tau) \theta_{2}\left(2\rho\vert 2\tau\right) \Bigg). \label{eq:almost_last}
\end{align}
where $Z^{D}_{0}(\tau), Z^{D}_{1/2}(\tau)$ are called the dual partition functions with integer and half-integer shifts (refer to \cite{Bonelli:2019boe} for further details on the partition functions).  The equation analogous to \eqref{eq:det_transcendent} is obtained by using the relation \eqref{id:theta1_23} in \eqref{proofeq:cb} and a straightforward simplification
\begin{gather}
    \frac{\theta_{3}\left( 2Q\vert 2\tau \right)}{\theta_{2}\left(  2Q\vert 2\tau \right)} = \frac{Z_{0}^{D}(\tau)}{Z_{1/2}^{D}(\tau)}.
\end{gather}

\subsection*{Further remarks}
\begin{itemize}
\item {\bf Relation to Lam\'e equation}

in a recent paper \cite{BGG} the authors show the following
\begin{itemize}
    \item[1.] the scalar form of the linear system \eqref{linear_system} at the zeros of the transcendent gives the Lam\'e equation,
    \item[2.] the accessory parameter of the Lam\'e equation is the Liouville action, i.e the difference of the action between the $\tau_{\star}$ where $Q(\tau_{\star}) =0$ and the asymptote $\tau \rightarrow i \infty$
\end{itemize}
further solidifying the Painlev\'e/CFT correspondence at the $c \rightarrow \infty$ limit. A rigorous proof of this semiclassical limit will be presented in an upcoming paper \cite{PCB2022}. 
    \item  {\bf Connection constant}

An immediate consequence of the Fredholm determinant representation for the tau-function is the computation of the connection constant, which is equivalent to studying the modular transformation of the toric conformal block. On the torus, the connection constant corresponds to the choice of pants decomposition along the A-cycle vs the B-cycle. In turn, this corresponds to the ratio of the tau function in the limits $\tau \rightarrow i \infty$ and $\tau \rightarrow 0$. In an upcoming paper \cite{Modular2022} we evaluate this constant.

\end{itemize}

\newpage

\printbibliography

\end{document}